\documentclass[]{aa} 
\usepackage{graphicx}
\usepackage{txfonts}
\usepackage{textcomp}

\usepackage{supertabular}

\usepackage{natbib}
\bibpunct{(}{)}{;}{a}{}{,} 

\usepackage[colorlinks=true,urlcolor=blue,linkcolor=black,citecolor=black]{hyperref}
\begin{document}
   \title{Extrasolar planets in stellar multiple systems}

   \author{T. Roell
          \inst{1}
          \and
	  R. Neuh\"auser\inst{1}
          \and
          A. Seifahrt\inst{1,2,3}
          \and
          M. Mugrauer\inst{1}
          }

   \institute{Astrophysical Institute and University Observatory Jena, 
              Schillerg\"a\ss chen 2, 07745 Jena, Germany\\
              \email{\href{mailto:troell@astro.uni-jena.de}{troell@astro.uni-jena.de}}
         \and
	      Physics Department, University of California, Davis, CA 95616, USA
         \and
	      Department of Astronomy and Astrophysics, University of Chicago, IL 60637, USA
             }

   \date{Received ...; accepted ...}

 
   \abstract
   {}
    {Analyzing exoplanets detected by radial velocity (RV) or transit observations, we determine the multiplicity of exoplanet host stars in order to study the influence of a stellar companion on the properties of planet candidates.}
    {Matching the host stars of exoplanet candidates detected by radial velocity or transit observations with online multiplicity catalogs in addition to a literature search, 57 exoplanet host stars are identified having a stellar companion.}
    {The resulting multiplicity rate of at least 12\,\% for exoplanet host stars is about four times smaller than the multiplicity of solar like stars in general. The mass and the number of planets in stellar multiple systems depend on the separation between their host star and its nearest stellar companion, e.g. the planetary mass decreases with an increasing stellar separation. We present an updated overview of exoplanet candidates in stellar multiple systems, including 15 new systems (compared to the latest summary from 2009).}
   {}


   \maketitle


\section{Introduction}

More than 700 extrasolar planet (exoplanet) candidates were discovered so far \citep[][www.exoplanet.eu]{schneider_EPE_2011}, but the knowledge of their properties is strongly affected by observational bias and selection effects. Taking the solar system as an archetype, the target lists of exoplanet search programs so far originally consist of mostly single and solar like stars (regarding the spectral type and age). But the first planet candidate detected by the RV technique was found around the primary of the close spectroscopic binary $\mathrm{\gamma\,Cep}$ \citep{campbell_gamma_ceph_1988, gamma_ceph_hatzes_2003, gamma_ceph_ralph_2007}, which demonstrates the existence of planets in binaries.

In the last years, imaging campaigns found stellar companions around several dozen exoplanet host stars formerly believed to be single stars \citep[see e.g.][and references therein]{raghavan_two_suns_sky_2006, mugi_HD125612_HD212301_2009}. Most of these exoplanet candidates are in the S-type orbit configuration (exoplanet surrounding one stellar component of a binary), while the orbit of a planet around both stellar binary components is called P-type orbit. Such circumbinary planets are detectable by measuring eclipse timing variations as done for NN\,Ser \citep{beuermann_NNserpentis_2010}, HW\,Vir \citep{lee_HWvir_2009}, DP\,Leo \citep{quian_dp_leo_2010}, HU\,Aqr \citep{qian_HU_Aqr_2011,hinse_HU_Aqu_2012}, and UZ\,For \citep{dai_UZ_For_2010, potter_UZ_For_2011}. Kepler-16\,(AB)b, Kepler-34\,(AB)b, and Kepler-35\,(AB)b are detected by measuring the transit lightcurve and eclipse timing variations \citep{kepler16_2011_science, kepler_34_35_nature_2012}, thus these are confirmed circumbinary planets.  Due to a different formation and evolution scenario for planets in a P-type orbit (compared to the more common S-type orbit), this paper only considers exoplanets found in a S-type orbit.

Multiplicity studies, as done by \citet{mugi_mnras_2007} or \citet{eggenberger_naco_2007}, are looking for stellar companions around exoplanet host stars by direct imaging. As summarized in \citet{mugi_HD125612_HD212301_2009}, these studies found 44 stellar companions around stars previously not known to be multiple, which results in a multiplicity rate of about 17\,\%, while \citet{raghavan_two_suns_sky_2006} found a host star multiplicity of about 23\,\%. The multiplicity rate of solar like stars\footnote{defined as all main-sequence stars with a spectral type from F6 to K3 within 25 parsec, see \citet{Raghavan_stellar_multiplicity_2010}} was determined by \citet{Raghavan_stellar_multiplicity_2010} to $\mathrm{(46\pm2)\,\%}$. \citet{duquennoy_solar_stars_multip_1991} measured the multiplicity of 164 nearby G-dwarfs (within 22\,pc) to 44\,\% (57\,\% considering incompleteness).


\section{Extrasolar planets in stellar multiple systems}

The Deuterium Burning Minimum Mass (DBMM) of $\mathrm{13\,M_{Jup}}$ is currently the most common criterion to distinguish a brown dwarf from a planet. However, we make use of the \textit{Extrasolar Planets Encyclopaedia} (hereafter EPE) in this paper and thus apply the definition of \citet{schneider_EPE_2011} who includes all confirmed substellar companions with a mass of less than $\mathrm{25\,M_{Jup}}$ within a $\mathrm{1\,\sigma}$ uncertainty. Due to a missing publication of the planet detection, the exoplanet candidates GJ\,433\,b, $\mathrm{\rho}$\,CrB\,b, 91\,Aqr\,b, $\mathrm{\nu}$\,Oph\,b\&c, $\mathrm{\tau}$\,Gem\,b, HD\,59686\,b, HD\,106515A\,b, HD\,20781\,b\&c, and HD\,196067\,b are not included in this paper. Also, the stellar binary HD\,176051, where \citet{phases_V_2010} detected the astrometric signal of an exoplanet around one of the two stellar components, is not included in this study: Because the planet was found by ground based astrometric observation (using an optical interferometer), this detection still need to be confirmed by other techniques and the final planetary mass depends on which of the stellar components is the host star.

   \begin{figure}
    \centering
      \includegraphics[width=6.0cm]{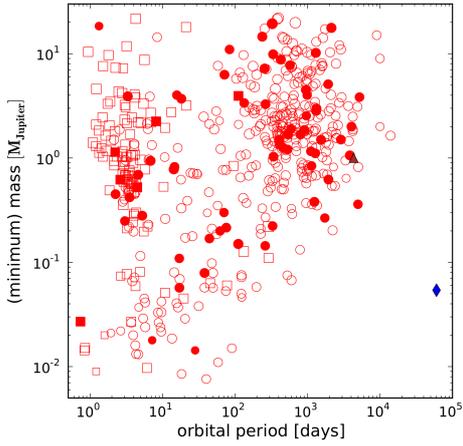}
    \caption{The (minimum) mass of extrasolar planets detected by RV (circles) or transit (squares) observations over their orbital period. Exoplanets around single stars are shown as open markers, while exoplanets in stellar multiple systems are coded by filled symbols. Jupiter is shown as a filled triangle and the filled diamond marks Neptune.}
    \label{fig:1}
   \end{figure}

The multiplicity of an exoplanet host star is defined (in this paper) by either a published common proper motion or an entry in the \textit{Catalogue of Components of Double and Multiple Stars} (hereafter CCDM) by \citet{ccdm_cat}. In case, the stellar multiplicity is only mentioned in the CCDM, all stellar components were checked on common proper motion using other catalogs (see appendix).
By searching the literature and matching the host stars of exoplanet candidates detected with transit or RV observations listed in the EPE (date: 2012/02/08) with the CCDM, 57 stellar multiple systems (47 double and 10 triple systems) with at least one exoplanet out of 477 systems in total are identified. The resulting multiplicity rate of about 12\,\% is less than previously published values (see table \ref{tab:multiplicity}). An explanation for that can be the increasing number of transiting exoplanets in the last years, which are included in this paper but excluded by previous studies. The host star multiplicity of transiting exoplanets is most likely still underestimated, because multiplicity studies around such host stars, like done by \citet{daemgen_multiplicity_transit_host_stars}, have just recently started.

\begin{table}[h]
\caption{Multiplicity of solar like and exoplanet host stars.}
\label{tab:multiplicity}
\centering
\begin{tabular}{c c c c c}\hline
Multiple	&	Single   &	Double	&	Triple or higher &	Reference \\ \hline
\multicolumn{5}{c}{\underline{solar like stars}}\\[0.2cm]
46\,\%		& 	54\,\% 	& 	34\,\%	&	9\,\%	&	1 \\
44\,\%		&	56\,\%	&	38\,\%	&	4\,\%	&	2 \\[0.1cm]
\multicolumn{5}{c}{\underline{exoplanet host stars}}\\[0.2cm]
22.9\,\%	&	77.1\,\%&	19.8\,\%&	3.1\,\%&	3 \\
17.2\,\%	&	82.8\,\%&	14.8\,\%&	2.4\,\%&	4 \\
11.95\,\% 	& 	88.05\,\%& 	9.85\,\%&	2.1\,\%	&	5 \\ \hline
\multicolumn{5}{l}{1: \citet{Raghavan_stellar_multiplicity_2010},\quad 2: \citet{duquennoy_solar_stars_multip_1991}}\\
\multicolumn{5}{l}{3: \citet{raghavan_two_suns_sky_2006},\quad 4: \citet{mugi_HD125612_HD212301_2009}}\\
\multicolumn{5}{l}{5: this work}
\end{tabular}
\end{table}

The complete list of the 57 multiple systems harboring exoplanet candidates can be found in the tables \ref{tab:exo_in_close_binaries}, \ref{tab:exo_in_wider_binaries}, and \ref{tab:exo_in_triples}. Furthermore, the proper motions of all these stars gathered from online catalogs are shown in the tables \ref{tab:systems_cpm}, \ref{tab:systems_diff_cpm}, and \ref{tab:systems_cpm_just_one_epoch}.
The latest published summary, done by \citet{mugi_HD125612_HD212301_2009}, listed 44 planetary systems in a stellar multiple system. However, two of these systems are excluded in this study, namely HD\,156846\,AB (after \citet{reffert_astrom_masses_2011} published astrometric mass limits of $\mathrm{m_{pl}=(10.5\ldots660.9)\,M_{Jup}}$, the EPE planetary status changed to unconfirmed), and 91\,Aqr (the planet detection itself is still not published in a refereed paper). In addition to that 42 systems, 15 new systems are listed and marked by the symbol \textcurrency\, in the tables \ref{tab:exo_in_close_binaries}, \ref{tab:exo_in_wider_binaries}, and \ref{tab:exo_in_triples}.

\begin{table*}
\centering
\caption[]{Critical \mbox{semi-major} axis $\mathrm{a_{crit}}$ for planets in close stellar binaries, calculated according to \citet{holman_exoplanet_stability_binaries_1999}. \\ \S\,... HD\,19994\,B itself is a close stellar binary with a total mass of $\mathrm{M_{BC}=0.9\,M_\odot}$ \citep{roell_paris_2011}. \\ \ddag\,... Values for eccentricity and \mbox{semi-major} axis of HD\,19994 are taken from \cite{eggenberger_Exo_Stat_III_2004}.}
\label{tab:2}
\begin{tabular}{lcccccccccc}\hline
Host star&$\mathrm{M_{host}}$&$\mathrm{M_{comp}}$&$\mathrm{\mu_{bin}}$&$\mathrm{e_{bin}}$ &$\mathrm{a_{bin}}$&$\mathrm{a_{crit}}$&$\mathrm{e_{pl}}$&$\mathrm{a_{pl}}$&$\mathrm{r^{apastron}_{pl}}$&References\\
 &$\mathrm{[M_\odot]}$&$\mathrm{[M_\odot]}$&&&[AU]&[AU]&&[AU]&[AU]& \\ \hline
$\mathrm{\gamma}$ Cep\,A & 1.40 & 0.41 & 0.23 & 0.41 & 20.2 & 3.86 & 0.05 & 2.05 & 2.15 &  \cite{gamma_ceph_ralph_2007}\\

HD\,41004\,A  & 0.70 & 0.42 & 0.38 & 0.40 & 20.0 & 3.38 & 0.39 & 1.60 & 2.28 &   \cite{chauvin_hd196885B_2011}\\

HD\,196885\,A & 1.33 & 0.45 & 0.25 & 0.42 & 21.0 & 3.84 & 0.48 & 2.60 & 3.85 &   \cite{chauvin_hd196885B_2011}\\

HD\,126614\,A&1.15&0.32&0.22&$\mathrm{\leq0.6}$&36.2&$\mathrm{\geq4.24}$&0.41&2.35&3.13&\cite{howard_Cal_Planet_search_2010} \\

HD\,19994\,A  & 1.34 & 0.90$^\S$ & 0.40 & 0.0$^\ddag$ &$\mathrm{\sim 100}^\ddag$& $\mathrm{\sim 31}$ & 0.30 &  1.42 & 1.85 &   \cite{roell_paris_2011, mayor_CORALIE_XII_2004}\\ \hline
\end{tabular}
\end{table*}


\section{Comparison of extrasolar planets in stellar multiple systems and around single stars}

\begin{figure}
    \centering
      \includegraphics[width=8.7cm]{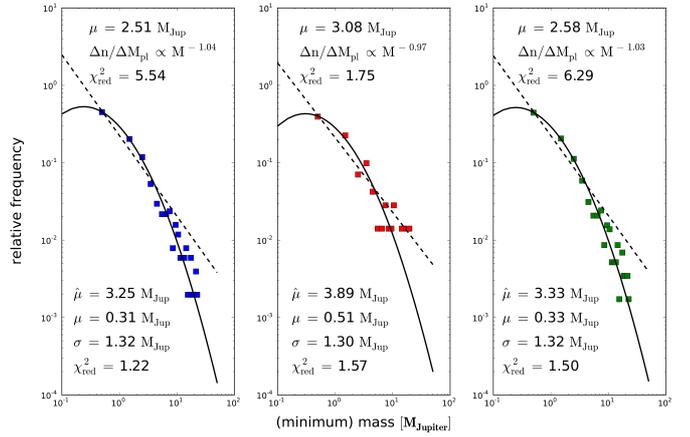}
    \caption{Mass distribution of exoplanets detected by RV or transit observations around single stars (left column), in stellar multiple systems (middle column), and for all kind of host stars (right column) fitted by a power law (dashed line, upper values) and a log-normal distribution (solid line, lower values)}
    \label{fig:4}
\end{figure}

\citet{marcy_props_exoplanets_2005} fitted the histogram of all known RV exoplanet minimum masses by a simple power-law and found an exponent of\, -1.05 for the mass distribution. That exponent is in good agreement with a sample of synthetic exoplanets detectable by current RV observations modeled by \citet{mordasini_exoplanet_pop_synt_II_2009}.
In our work, planets currently found by RV or transit observations are analyzed (see Fig. \ref{fig:1}). Using also a simple power-law (see Fig. \ref{fig:4}), an exponent of\, -1.03 was found for the mass distribution of all exoplanet candidates, which is similar to the results of previous works. For exoplanet candidates in stellar multiple systems and around single stars, the exponent is {-0.97} and {-1.04}, respectively. The mean of the planetary masses is about $\mathrm{2.5\,M_{Jup}}$ for planets around single stars and $\mathrm{3.1\,M_{Jup}}$ for the case of stellar multiplicity.
In addition to the power-law, we also fit a log-normal distribution to the planetary masses. The probability distribution function (PDF) and the expectation value $\mathrm{\hat{\mu}}$ of a log-normal distribution for a measure x can be calculated by
\begin{equation}
 \mathrm{PDF\,(x,\mu,\sigma) = e^{-\frac{(\ln{x}-\mu)^2}{2\,\sigma^2}}\;/\;(x\,\sigma\,\sqrt{2\,\pi})},\quad \mathrm{\hat{\mu} = e^{\,\mu\,+\,\sigma^2/2}}
\label{equ:PDF_Log_Normal}
\end{equation}
where $\mathrm{\mu}$ and $\mathrm{\sigma}$ are the mean value and the standard deviation of the distribution. To determine the $\mathrm{\chi^2}$ value (shown in Fig. \ref{fig:4}) the Python package ``SciPy'' \citep{python_scipy} was used.

As one can see in Fig \ref{fig:4}, the log-normal fit results in a better $\chi^2_{red}$ than the power-law fit. The expectation values for the mass of exoplanets around single stars and in stellar multiple systems differ, hence the power-law as well as the log-normal fit lead to the conclusion that the mass distribution of exoplanets in stellar multiple systems are pushed towards higher planetary masses, compared to the mass distribution of exoplanets around single stars.

However, the statistic of the exoplanet host star multiplicity is still affected by observational bias and selection effects of the originally planet search programs. Most multiplicity studies so far were carried out after the planet detection. Hence, most of the host stars are solar like (regarding the age and spectral type), but they are also originally selected as single stars. To avoid the adaption of such selection effects, systematic searches for planets in stellar multiple systems, like described in \citet{desidera_sarg_2007} or \citet{roell_torun_2010}, are needed.

%

\section{Influence of a close stellar companion on planet properties}

\begin{figure}[b]
    \centering
      \includegraphics[width=6cm]{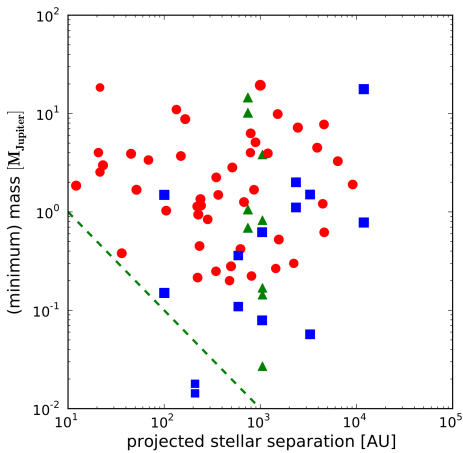}
    \caption{Planetary (minimum) mass over the projected separation of the exoplanet host star and its nearest stellar companion. The markers represent the number of planets per system (dots ... one planet, squares ... two planets, triangles ... three or more planets).  The size of the symbols represent the mass of the exoplanet host star.}
    \label{fig:5}
\end{figure}

In the previous section, the difference in the mass of exoplanets around single stars and in stellar multiple systems was discussed. In order to unveil the cause of this difference, a closer look on the influence of a stellar companion around the exoplanet host star is advisable. In Fig. \ref{fig:5} we plot the planetary (minimum) mass over the projected separation of the exoplanet host star and its nearest stellar companion. Because all systems analyzed in this paper are hierarchical, the exoplanet host star and its nearest stellar companion can be treated as a binary system. The order of the stellar multiplicity is not relevant, but the planetary minimum mass decreases with an increasing projected stellar separation (dashed line in Fig. \ref{fig:5}). Furthermore, multi-planet systems are only present in stellar systems with a projected stellar separation larger than about 100\,AU and up to now, no planet was found in a stellar binary with a projected separation of less than 10\,AU.
The two planets below the dashed line in Fig. \ref{fig:5} are the planetary system around GJ\,667\,C, a component of a hierarchical triple star system at a distance of 7\,pc. The true semi-major axis is likely larger than the measured projected separation and the true planetary mass could also be larger than the measured minimum mass. These observational bias effects could explain, why GJ\,667\,C is the only system left of that dashed line in Fig. \ref{fig:5}.

\begin{figure}
    \centering
      \includegraphics[width=6cm]{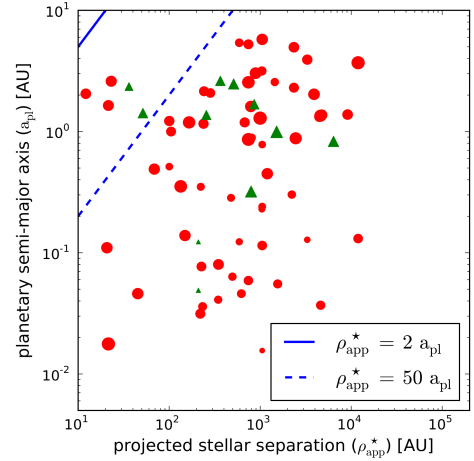}
    \caption{Planetary semi-major axis over the projected stellar separation (dots ... stellar binary, triangles ... triple star). The five systems in the upper left ($\mathrm{\rho^\star_{app} < 50\,a_{pl}}$) are shown in more detail in table \ref{tab:2}. The size of the symbols represent the (minimum) mass of the exoplanet.}
    \label{fig:6}
\end{figure}

\citet{holman_exoplanet_stability_binaries_1999} determine a formula to calculate the critical \mbox{semi-major} axis $\mathrm{a_{crit}}$ for a stable planetary orbit coplanar to the stellar orbit with the \mbox{semi-major} axis $\mathrm{a_{bin}}$, which varies from $\mathrm{a_{crit} \simeq (0.02\ldots0.45)\,a_{bin}}$, depending on the mass ratio $\mathrm{\mu_{bin}}$ and the eccentricity $\mathrm{e_{bin}}$ of the stellar binary. Table \ref{tab:2} listed the five systems, where the apparent separation is less than 50 times the planetary semi-major axis (see Fig. \ref{fig:6}) including the corresponding critical \mbox{semi-major} axis. Except for the exoplanet HD\,196885\,Ab, which grazes an ``unstable region'' during the apastron passage, all these systems are clearly stable. However, considering the age of the F8V star HD\,196885\,A of $\mathrm{2.0\pm0.5\,Gyr}$ \citep{correia_elodie_IV_2008}, the planetary system can also be regarded as long-term stable.

%
\section{Summary}

Analyzing the host star multiplicty of exoplanets detected by RV or transit observations, 57 exoplanet host stars with stellar companions are identified and presented in the appendix, including 15 new systems (compared to the latest published summary in 2009).
The resulting multiplicity rate for exoplanet host stars of at least 12\,\% is about four times smaller than the multiplicity of solar like stars. No planet is found so far in stellar binaries with a projected separation of less than 10\,AU and multi-planet systems were only found in stellar systems with a projected separation larger than 100\,AU. The planetary (minimum) mass decreases with an increasing projected stelllar separation.

%

\begin{acknowledgements}
      T.R. and R.N. thank the DFG
      (\emph{Deut\-sche For\-schungs\-ge\-mein\-schaft}) for financial support under the project numbers NE\,515/23-1, NE\,515/30-1, and NE\,515/33-1 (SPP\,1385: ``First ten million years of the solar systems''). A.S. acknowledge support from the National Science Foundation under grant NSF\,AST-0708074. This research has made use of the SIMBAD database and the VizieR catalog access tool, both operated at CDS, Strasbourg, France.
\end{acknowledgements}

\bibliographystyle{aa}
\bibliography{literatur_datenbank}

\begin{appendix}
\onecolumn
\section{Updated tables of extrasolar planets in stellar multiple systems}

\vspace{1.5cm}
\begin{center}

\tablefirsthead{
CCDM $\ldots$		& \multicolumn{5}{l}{\textit{Catalog of Components of Double and Multiple stars}, \cite{ccdm_cat}}	\\
ASCC-2.5 V3 $\ldots$	& \multicolumn{5}{l}{\textit{All-sky Compiled Catalogue of 2.5 million stars (3rd version, 2009)}, \cite{ASCC_2.5_v3}}	\\
HIP-2 $\ldots$		& \multicolumn{5}{l}{\textit{Hipparcos, the New Reduction}, \cite{new_HIP_2008}}	\\
Nomad-1 $\ldots$	& \multicolumn{5}{l}{\textit{Naval Observatory Merged Astrometric Dataset}, \cite{NOMAD}}	\\
UCAC-3 $\ldots$		& \multicolumn{5}{l}{\textit{Third U.S. Naval Observatory CCD Astrograph Catalog}, \cite{UCAC3}}\\
PPMXL $\ldots$		& \multicolumn{5}{l}{\textit{The PPMXL catalog of positions and proper motions on the ICRS. Combining USNO-B1.0}}\\
			& \multicolumn{5}{l}{\textit{and the two Micron All Sky Survey (2MASS).}, \cite{PPMXL}}\\
Tycho-2 $\ldots$	& \multicolumn{5}{l}{\textit{The Tycho-2 Catalogue of the 2.5 Million Brightest Stars}, \cite{TYCHO2}}\\[1.5cm]
\hline System & EPE planet status	& r [''] & $\mathrm{\mu_\alpha \cos{\delta}\,[mas/year]}$ & $\mathrm{\mu_\delta\,[mas/year]}$ & Catalogue\\ \hline	& & & & & \\
}

\tablehead{\hline System & EPE planet status	& r [''] & $\mathrm{\mu_\alpha \cos{\delta}\,[mas/year]}$ & $\mathrm{\mu_\delta\,[mas/year]}$ & Catalogue\\ \hline	& & & & & \\}

\tabletail{\hline \multicolumn{6}{r}{\textit{continued on next page}}\\ \hline}
\tablelasttail{ \hline}

\topcaption{Exoplanet host stars having a common proper motion with another star. The ID of the host stars is the same as it is used in the EPE and the planet status is coded by $\mathrm{R\ldots}$ published in a \textbf{R}efereed paper, $\mathrm{S\ldots}$ \textbf{S}ubmitted to a professional journal, and $\mathrm{C\ldots}$ announced by astronomers in professional \textbf{C}onferences. A planet status followed by the letter R in brackets means, the status on the EPE is not up-to-date and the planet detection is already published in a refereed paper. The fourth and fifth column show the proper motion listed in the catalog mentioned in the last column. For easy identification of the stellar companions in the online catalogs, the third column contains either the separation from the primary as calculated by the used online catalog or the latest separation measurement in case of published relative astrometric measurements.}

\label{tab:systems_cpm}

\begin{supertabular}{r c c r r r}

\object{11 Com}\,A 	& R	&	9	& $-109.37\pm1.26$	& $89.25\pm0.75$ 	&	ASCC-2.5-V3	\\
\object{11 Com}\,B 	& -	&	-	& $-109\pm$ n.s.	& $85\pm$ n.s. 		&	CCDM	\\[0.2cm]

\object{16 Cyg}\,A 	& -	&	-	& $-133.39\pm0.82$	& $-163.08\pm0.85$ 	&	ASCC-2.5-V3	\\
\object{16 Cyg}\,B 	& R	&	39.44	& $-147.77\pm0.85$	& $-158.39\pm0.76$	&	ASCC-2.5-V3	\\
\object{16 Cyg}\,C 	& -	&\multicolumn{4}{l}{close C component about 3.4'' away from A, see \cite{patience_2002}}	\\[0.2cm]

\object{30 Ari}\,A 	& -	&	-	& $137.66\pm0.74$ 	& $-14.98\pm0.91$ 	&	ASCC-2.5-V3	\\
			& -	&\multicolumn{4}{l}{A itself is a spectroscopic binary, see \cite{eike_30_Ari_2009}}	\\
\object{30 Ari}\,B 	& R	&	36.86	& $145.33\pm1.21$	& $-12.86\pm0.94$ 	&	ASCC-2.5-V3	\\[0.2cm]

\object{55 Cnc}\,A 	& R	&	-	& $-485.4\pm0.9$ 	& $-234.4\pm0.7$	&	Nomad-1	\\
\object{55 Cnc}\,B 	& -	&	84.16	& $-488.0\pm6.0$	& $-234.0\pm5.0$ 	&	Nomad-1	\\[0.2cm]

\object{91 Aqr}\,A 	& S	&	-	& $370.70\pm0.63$	& $-16.22\pm0.7$ 	&	ASCC-2.5 V3	\\
\object{91 Aqr}\,B 	& -	&	51.93	& $373.00\pm2.41$	& $-18.46\pm2.41$ 	&	ASCC-2.5 V3	\\
\object{91 Aqr}\,C 	& -	&	52.03	& $370.00\pm1.87$	& $-20.34\pm0.98$ 	&	ASCC-2.5 V3	\\
\object{91 Aqr}\,D 	& -	&	106.67	& $18.3\pm8.4$		& $-10.5\pm7.8$ 	&	Nomad-1		\\
\object{91 Aqr}\,E 	& -	&	88.05	& $-9.4\pm8.7$		& $-25.2\pm7.8$ 	&	Nomad-1		\\[0.2cm]

\object{$\mathrm{\gamma}$ Cep}\,A & R	&	-	& $-48.8\pm0.4$ & $127.1\pm0.4$ &	Nomad-1	\\
\object{$\mathrm{\gamma}$ Cep}\,B & -	&	0.02	& \multicolumn{3}{l}{imaged directly by \cite{gamma_ceph_ralph_2007}}\\[0.2cm]

\object{$\mathrm{\gamma^1}$ Leo}\,A 	& R	&	-	& $306.35\pm3.27$ 	& $-160.77\pm2.30$ 	&	ASCC-2.5 V3	\\
\object{$\mathrm{\gamma^1}$ Leo}\,B 	& -	&	4.33	& $309.60\pm1.32$ 	& $-152.91\pm0.75$ 	&	ASCC-2.5 V3	\\
\object{$\mathrm{\gamma^1}$ Leo}\,C 	& -	&	325.15	& $-501.30\pm1.05$ 	& $-41.97\pm1.28$ 	&	ASCC-2.5 V3	\\
\object{$\mathrm{\gamma^1}$ Leo}\,D 	& -	&	366.86	& $-10.01\pm1.53$ 	& $-22.85\pm1.21$ 	&	ASCC-2.5 V3	\\[0.2cm]

\object{$\mathrm{\tau}$ Boo}\,A & R	&	-	& $-479.53\pm0.16$ & $53.49\pm0.13$ 	&	HIP-2	\\
\object{$\mathrm{\tau}$ Boo}\,B & -	&	2.8	& \multicolumn{3}{l}{confirmed by \cite{duquennoy_solar_stars_multip_1991}}\\[0.2cm]

\object{$\mathrm{\upsilon}$ And}\,A 	& R	&	-	& $-172.5\pm0.5$ 	& $-381.0\pm0.4$ &	Nomad-1	\\
\object{$\mathrm{\upsilon}$ And}\,B 	& -	&	55	&\multicolumn{3}{l}{confirmed by \cite{lowrance_ups_and_B_2002}}\\
\object{$\mathrm{\upsilon}$ And}\,C 	& -	&	110.27	& $-9.6\pm2.3$ 		& $-3.5\pm2.$ 	&	Nomad-1	\\
\object{$\mathrm{\upsilon}$ And}\,D 	& -	&	273.20	& $16.3\pm0.7$ 		& $-4.7\pm0.6$ 	&	Nomad-1	\\[0.2cm]

\object{GJ 667 AB} 	& -	&	-	& $1129.76\pm9.72$	& $-77.02\pm4.67$	&	HIP-2	\\
			& -	&	-	& $1171.69\pm2.70$	& $-168.80\pm2.79$	&	ASCC-2.5 V3	\\
			& -	&	-	& $1161.4\pm2.3$	& $-172.3\pm2.3$	&	PPMXL	\\
\object{GJ 667}\,C 	& R	&	32.66	& $1049.0\pm$ n.s.	& $-91.0\pm$ n.s.	&	UCAC-3	\\
			& 	&	32.75	& $1155.0\pm7.2$	& $-214.4\pm9.5$	&	NOMAD-1	\\
\multicolumn{6}{l}{GJ\,667\,C has not a common proper motion within the measurement errors, but it can be ruled out as a background object.}\\
\multicolumn{6}{l}{The large spread in the proper motion measurements can be possibly explained by the fact, that the AB component}\\
\multicolumn{6}{l}{influences the relative astrometric measurements of the (AB)\&C pair.}\\[0.2cm]

\object{GJ 676}\,A 	& R	&	-	& $-259.2\pm1.4$	& $-185.6\pm0.9$ 	&	Nomad-1	\\
\object{GJ 676}\,B 	& -	&	48.98	& $-254.0\pm8.0$	& $-156.0\pm23.0$ 	&	Nomad-1	\\
			& 	&	47.78	& $-294.6\pm7.4$	& $-186.8\pm7.0$ 	&	UCAC-3	\\[0.2cm]

\object{GJ 3021}\,A 	& R	& 	-	& $433.8\pm0.5$		& $-57.9\pm0.4$ 	&	Nomad-1	\\
\object{GJ 3021}\,B 	& -	&	3.86	& \multicolumn{3}{l}{confirmed by \cite{chauvin_2006}}	\\[0.2cm]

\object{Gl 86}\,A 	& R	&	-	& $2183.43\pm3.20$ 	& $661.70\pm3.50$ 	&	ASCC-2.5-V3	\\
\object{Gl 86}\,B 	& -	&	1.93	& \multicolumn{3}{l}{confirmed by \cite{mugi_gl86B_2005}}	\\[0.2cm]

\object{HAT-P-1}\,A 	& -	&	-	& $30.0\pm0.6$ 		& $-42.3\pm1.3$ 	&	UCAC-3	\\
\object{HAT-P-1}\,B 	& R	&	10.93	& $32.6\pm0.8$ 		& $-43.2\pm1.9$ 	&	UCAC-3	\\[0.2cm]

\object{WASP-8}\,A 	& S(R)	&	-	& $109.62\pm2.05$ 	& $10.02\pm1.46$ 	&	ASCC-2.5 V3	\\
\object{WASP-8}\,B 	& -	&	 	& \multicolumn{3}{l}{confirmed by \cite{queloz_wasp8_2010}}	\\
\object{WASP-8}\,C 	& -	&	140.93	& $50.33\pm1.39$ 	& $-28.65\pm1.08$ 	&	ASCC-2.5 V3	\\[0.2cm]

\object{XO-2}\,A 	& C(R)	&	-	& $-34.72\pm2.50$ 	& $-153.61\pm2.40$ 	&	ASCC-2.5 V3	\\
\object{XO-2}\,B 	& -	&	30.05	& $-33.02\pm2.89$ 	& $-154.11\pm2.70$ 	&	ASCC-2.5 V3	\\[0.2cm]

\object{HD 142}\,A 	& R	&	-	& $575.2\pm0.4$		& $-39.9\pm0.5$ 	&	Nomad-1	\\
\object{HD 142}\,B 	& -	&	4.10	& \multicolumn{3}{l}{confirmed by \cite{eggenberger_naco_2007}}\\[0.2cm]

\object{HD 3651}\,A 	& R	&	-	& $-460.22\pm0.89$ 	& $-370.22\pm0.75$ 	&	ASCC-2.5 V3	\\
\object{HD 3651}\,B 	& -	&	43.07	& \multicolumn{3}{l}{confirmed by \cite{mugi_HD3651_2006}}	\\
\object{HD 3651}\,C 	& -	&	168.26	& $10.81\pm2.00$ 	& $-1.48\pm2.40$ 	&	ASCC-2.5 V3	\\[0.2cm]

\object{HD 7449}\,A 	& R	&	-	& $-161.6\pm0.6$ 	& $-138.9\pm0.5$ 	&	PPMXL	\\
\object{HD 7449}\,B 	& 	&	60.20	& $-162.5\pm6.1$ 	& $-137.5\pm6.1$ 	&	PPMXL	\\[0.2cm]

\object{HD 16141}\,A 	& R	& 	-	& $-155.33\pm1.19$	& $-437.42\pm1.23$ 	&	ASCC-2.5 V3	\\
\object{HD 16141}\,B 	& -	&	6.20	& \multicolumn{3}{l}{confirmed by \cite{mugi_2005}}	\\[0.2cm]

\object{HD 11964}\,A 	& R	&	-	& $-366.10\pm0.3$	& $-240.83\pm0.62$ 	&	ASCC-2.5 V3	\\
\object{HD 11964}\,B 	& -	&	30.68	& $-369.98\pm4.34$	& $-245.47\pm3.13$ 	&	ASCC-2.5 V3	\\[0.2cm]


\object{HD 19994}\,A 	& R	&	-	& $194.56\pm0.37$ 	& $-69.01\pm0.30$ &	HIP-2	\\
\object{HD 19994 BC} 	& -	&	2.30	& \multicolumn{3}{l}{confirmed by \cite{duquennoy_solar_stars_multip_1991}}\\
			& 	&\multicolumn{4}{l}{close C component around B found by \cite{roell_paris_2011}}\\[0.2cm]

\object{HD 20782}\,A 	& R	&	-	& $349.85\pm0.80$	& $-65.00\pm1.07$ 	&	ASCC-2.5 V3	\\
\object{HD 20782}\,B 	& -	&	253.71	& $349.76\pm1.13$ 	& $-68.44\pm1.57$ 	&	ASCC-2.5 V3	\\[0.2cm]

\object{HD 27442}\,A 	& R	&	-	& $-48.09\pm0.80$ 	& $-166.69\pm0.77$ 	&	ASCC-2.5 V3	\\
\object{HD 27442}\,B 	& -	&	13.06	& \multicolumn{3}{l}{confirmed by \cite{chauvin_2006}}\\[0.2cm]

\object{HD 28254}\,A 	& S(R)	&	-	& $-66.9\pm0.6$ 	& $-144.0\pm0.6$ &	Nomad-1	\\
\object{HD 28254}\,B 	& -	&	4.3	& \multicolumn{3}{l}{confirmed by \cite{naef_harps_XXIII_2010}}	\\[0.2cm]

\object{HD 38529}\,A 	& R	& 	-	& $-79.22\pm1.02$	& $-142.22\pm1.00$ 	&	ASCC-2.5 V3	\\
\object{HD 38529}\,B 	& -	&	283.72	& $-81.79\pm14.50$	& $-117.47\pm14.80$ 	&	ASCC-2.5 V3	\\
			& 	&	\multicolumn{4}{l}{B component confirmed by \cite{raghavan_two_suns_sky_2006}}	\\[0.2cm]

\object{HD 40979}\,A 	& R	& 	-	& $96.40\pm1.19$	& $-153.36\pm0.70$ 	&	ASCC-2.5 V3	\\
\object{HD 40979 BC} 	& -	&	193.77	& $94.08\pm1.85$ 	& $-153.11\pm1.65$ 	&	ASCC-2.5 V3	\\
			& -	&\multicolumn{4}{l}{C component about 3.9'' away from B found by \cite{mugi_GJ3021_HD27442_2007}}\\[0.2cm]

\object{HD 41004}\,A 	& R	&	-	& $-41.72\pm1.19$ 	& $64.87\pm1.34$ 	&	ASCC-2.5 V3	\\
\object{HD 41004}\,B 	& R	&	1.19	& $-42.25\pm1.08$ 	& $65.16\pm1.12$ 	&	ASCC-2.5 V3	\\[0.2cm]

\object{HD 46375}\,A 	& R	&	-	& $114.2\pm0.9$ 	& $-96.7\pm0.7$ 	&	Nomad-1	\\
\object{HD 46375}\,B 	& 	&	10.35	& \multicolumn{3}{l}{confirmed by \cite{mugi_calar_alto_survey_2006}}	\\[0.2cm]

\object{HD 65216}\,A 	& R	&	-	& $-122.66\pm1.34$ 	& $145.77\pm1.15$ 	&	ASCC-2.5 V3	\\
\object{HD 65216 BC} 	& -	&	7.14	& \multicolumn{3}{l}{confirmed by \cite{mugi_mnras_2007}}	\\
			& 	&\multicolumn{4}{l}{C component 0.17'' away from B found by \cite{mugi_mnras_2007}}\\[0.2cm]

\object{HD 75289}\,A 	& R	& 	-	& $-19.90\pm0.66$	& $-228.13\pm0.73$ 	&	ASCC-2.5 V3	\\
\object{HD 75289}\,B 	& -	&	21.47	& \multicolumn{3}{l}{confirmed by \cite{mugi_HD75289_2004}}	\\[0.2cm]

\object{HD 80606}\,A 	& R	&	-	& $56.84\pm1.37$ 	& $10.02\pm1.71$ 	&	ASCC-2.5 V3	\\
\object{HD 80606}\,B 	& 	&	20.18	& $51.95\pm1.32$ 	& $9.97\pm1.77$ 	&	ASCC-2.5 V3	\\[0.2cm]

\object{HD 89744}\,A 	& R	& 	-	& $-120.12\pm0.85$	& $-138.66\pm0.76$ 	&	ASCC-2.5 V3	\\
\object{HD 89744}\,B 	& -	&	62.99	& \multicolumn{3}{l}{confirmed by \cite{mugi_HD89744_2004}}	\\[0.2cm]

\object{HD 99491}\,(A)	& -	&	-	& $-725.22\pm0.64$ 	& $180.30\pm0.72$ 	&	ASCC-2.5 V3	\\
\object{HD 99492}\,(B)	& R	&	33.27	& $-727.57\pm1.55$ 	& $186.47\pm1.55$ 	&	ASCC-2.5 V3	\\
\object{HD 99492}\,(C)	& -	&	204.56	& $5.36\pm1.98$ 	& $-13.72\pm2.00$ 	&	ASCC-2.5 V3	\\[0.2cm]

\object{HD 101930}\,A 	& R	& 	-	& $15.42\pm1.30$	& $348.29\pm1.27$ 	&	ASCC-2.5 V3	\\
\object{HD 101930}\,B 	& -	&	69.89	& $24.98\pm4.01$ 	& $351.54\pm2.35$ 	&	ASCC-2.5 V3	\\
			& -	&	\multicolumn{4}{l}{B component confirmed by \cite{mugi_mnras_2007}}	\\[0.2cm]

\object{HD 109749}\,A 	& C(R)	&	-	& $-156.69\pm1.47$	& $-4.88\pm1.54$ 	&	ASCC-2.5 V3	\\
\object{HD 109749}\,B 	& -	&	8.42	& $-157.88\pm1.42$	& $-5.46\pm1.28$ 	&	ASCC-2.5 V3	\\[0.2cm]

\object{HD 114729}\,A 	& R	& 	-	& $-200.83\pm1.23$	& $-307.82\pm1.11$ 	&	ASCC-2.5 V3	\\
\object{HD 114729}\,B 	& -	&	8.05	& \multicolumn{3}{l}{confirmed by \cite{mugi_2005}}	\\[0.2cm]

\object{HD 114762}\,A 	& R	& 	-	& $-582.75\pm1.12$	& $-1.04\pm1.07$ 	&	ASCC-2.5 V3	\\
\object{HD 114762}\,B 	& -	&	3.26	& \multicolumn{3}{l}{confirmed by \cite{patience_2002}}	\\[0.2cm]

\object{HD 125612}\,A 	& S(R)	& 	-	& $-62.25\pm1.66$	& $-67.63\pm1.50$ 	&	ASCC-2.5 V3	\\
\object{HD 125612}\,B 	& -	&	89.99	& \multicolumn{3}{l}{confirmed by \cite{mugi_HD125612_HD212301_2009}}	\\[0.2cm]

\object{HD 126614 AB}	& S(R)	&	-	& $-151.66\pm1.35$	& $-148.66\pm1.22$ 	&	ASCC-2.5 V3	\\
			& 	&	-	& $-152.4\pm1.0$	& $-148.1\pm0.9$ 	&	Nomad-1	\\
			\multicolumn{6}{r}{B is 0.5'' away from A, found by \cite{howard_Cal_Planet_search_2010}}\\
			\multicolumn{6}{r}{and common proper motion confirmed by \cite{ginski_mnras_2012}}\\[0.1cm]
\object{HD 126614}\,C 	& -	&	41.80	& $-144.0\pm4.0$ 	& $-142.0\pm3.0$ 	&	Nomad-1	\\[0.2cm]

\object{HD 132563}\,A 	& -	&	-	& $-56.38\pm2.22$	& $-68.87\pm1.69$ 	&	ASCC-2.5 V3	\\
			& 	&\multicolumn{4}{l}{A itself is a spectroscopic binary, see \cite{desidera_hd132563_2011}}\\
\object{HD 132563}\,B 	& R	&	3.79	& $-57.15\pm2.11$	& $-69.18\pm1.62$ 	&	ASCC-2.5 V3	\\
\object{HD 132563}\,C 	& -	&	64.77	& $-14.52\pm1.57$	& $7.36\pm1.02$ 	&	ASCC-2.5 V3	\\[0.2cm]

\object{HD 137388}\,A 	& R	&	-	& $-46.72\pm0.89$	& $43.31\pm0.97$ 	&	HIP-2	\\
\object{HD 137388}\,B 	& -	&	21.31	& $-35.54\pm14.33$	& $74.43\pm16.63$ 	&	HIP-2	\\
			& -	&	21.32	& $-94.69\pm58.77$	& $38.04\pm63.18$ 	&	ASCC-2.5 V3	\\[0.2cm]

\object{HD 142022}\,A 	& R	&	-	& $-340.52\pm0.91$	& $-31.04\pm1.01$ 	&	ASCC-2.5 V3	\\
\object{HD 142022}\,B 	& -	&	22.85	& $-359.70\pm2.38$	& $-26.28\pm2.04$ 	&	ASCC-2.5 V3	\\[0.2cm]

\object{HD 147513}\,A 	& R	& 	-	& $72.47\pm0.93$	& $3.43\pm0.83$ 	&	ASCC-2.5 V3	\\
\object{HD 147513}\,B 	& -	&	345.69	& $76.73\pm2.17$	& $1.94\pm2.10$ 	&	ASCC-2.5 V3	\\[0.2cm]

\object{HD 177830}\,A 	& R	& 	-	& $-40.83\pm0.75$	& $-50.13\pm0.99$ 	&	ASCC-2.5 V3	\\
\object{HD 177830}\,B 	& -	&	1.65	& \multicolumn{3}{l}{confirmed by \cite{eggenberger_naco_2007}}	\\[1.2cm]

\object{HD 178911}\,A 	& -	&	-	& $50.20\pm1.87$ 	& $190.05\pm2.46$	&	ASCC-2.5 V3	\\
			& 	&	-	& $51.89\pm2.09$ 	& $196.24\pm2.31$ 	&	HIP-2	\\
			& -	&\multicolumn{4}{l}{A itself is a spectroscopic binary, see \cite{McAlister_speckle_binaries_II_1987}}\\
\object{HD 178911}\,B 	& R	&	16.86	& $67.13\pm2.11$ 	& $190.50\pm2.64$ 	&	ASCC-2.5 V3	\\
			& 	&	16.85	& $55.14\pm3.43$ 	& $201.30\pm4.39$ 	&	HIP-2	\\[0.2cm]

\object{HD 185269}\,A	& R	&	-	& $-32.3\pm0.5$ 	& $-80.7\pm0.6$		&	PPMXL	\\
\object{HD 185269}\,B	& -	&	4.51	& \multicolumn{3}{l}{confirmed by \cite{ginski_mnras_2012}}\\[0.2cm]

\object{HD 188015}\,A 	& R	& 	-	& $55.61\pm0.12$	& $-91.62\pm1.11$ 	&	ASCC-2.5 V3	\\
\object{HD 188015}\,B 	& -	&	13	& \multicolumn{3}{l}{confirmed by \cite{raghavan_two_suns_sky_2006}}	\\[0.2cm]

\object{HD 189733}\,A 	& R	& 	-	& $-2.39\pm0.87$	& $-250.19\pm0.80$ 	&	ASCC-2.5 V3	\\
\object{HD 189733}\,B 	& -	&	11.37	&\multicolumn{3}{l}{confirmed by \cite{bakos_HD189733B_2006}}	\\[0.2cm]

\object{HD 190360}\,A 	& R	&	-	& $683.3\pm0.4$ 	& $-524.0\pm0.5$ 	&	Nomad-1	\\
\object{HD 190360}\,B 	& -	&	178.00	& $686.8\pm4.1$ 	& $-530.3\pm4.1$ 	&	Nomad-1	\\[0.2cm]

\object{HD 195019}\,A 	& R	&	-	& $349.54\pm1.12$ 	& $-57.27\pm0.82$ 	&	ASCC-2.5 V3	\\
\object{HD 195019}\,B 	& -	&\multicolumn{4}{l}{Multiplicity first mentioned by \cite{fischer_hd195019_hd217107_1999}}\\[0.2cm]

\object{HD 196050}\,A 	& R	& 	-	& $-191.30\pm0.85$	& $-63.31\pm0.99$ 	&	ASCC-2.5 V3	\\
\object{HD 196050 BC} & -	&	10.88	& \multicolumn{3}{l}{confirmed by \cite{mugi_2005}}	\\
			& -	&\multicolumn{4}{l}{C component about 0.4'' away from B found by \cite{eggenberger_naco_2007}}\\[0.2cm]

\object{HD 196885}\,A 	& R	&	-	& $47.4\pm0.8$ 		& $83.0\pm0.5$ 	&	Nomad-1	\\
			& 	&	-	& $56.5\pm1.1$ 		& $87.3\pm1.2$ 	&	Tycho-2	\\
\object{HD 196885}\,B 	& -	&	0.70	& \multicolumn{3}{l}{confirmed by \cite{chauvin_hd196885B_2011}}\\
\object{HD 196885}\,C 	& -	&	183.06	& $-5.2\pm1.4$ 		& $-3.9\pm1.3$ 	&	Nomad-1	\\
			& -	&	183.12	& $-3.1\pm1.5$ 		& $-2.2\pm1.4$ 	&	Tycho-2	\\[0.2cm]

\object{HD 204941}\,A 	& R	&	-	& $-298.24\pm1.34$ 	& $-124.68\pm0.67$ 	&	ASCC-2.5 V3	\\
\object{HD 204941}\,B 	& -	&	53.62	& $-308.85\pm3.48$ 	& $-124.13\pm1.84$ 	&	ASCC-2.5 V3	\\[0.2cm]

\object{HD 212301}\,A 	& S(R)	& 	-	& $79.12\pm1.04$	& $-92.37\pm0.99$ 	&	ASCC-2.5 V3	\\
\object{HD 212301}\,B 	& -	&	4.43	& \multicolumn{3}{l}{confirmed by \cite{mugi_HD125612_HD212301_2009}}	\\[0.2cm]

\object{HD 213240}\,A 	& R	&	-	& $-135.1\pm0.6$ 	& $-194.0\pm0.4$ 	&	Nomad-1	\\
\object{HD 213240}\,B 	& -	&	21.94	& $65.1\pm5.0$ 		& $-10.8\pm1.5$ 	&	Nomad-1	\\
\object{HD 213240}\,C 	& -	&	95.69	& \multicolumn{3}{l}{confirmed by \cite{mugi_2005}}	\\[0.2cm]

\object{HD 222582}\,A 	& R	&	-	& $-145.4\pm1.2$ 	& $-111.0\pm0.8$ 	&	Nomad-1	\\
\object{HD 222582}\,B 	& -	&	109.42	& $-147.5\pm4.4$ 	& $-114.2\pm4.4$ 	&	Nomad-1	\\

\end{supertabular}

\vspace{1.5cm}

\tablefirsthead{\hline System & EPE planet status	& r [''] & $\mathrm{\mu_\alpha \cos{\delta}\,[mas/year]}$ & $\mathrm{\mu_\delta\,[mas/year]}$ & Catalogue\\ \hline	& & & & & \\}

\tablehead{\hline System & EPE planet status	& r [''] & $\mathrm{\mu_\alpha \cos{\delta}\,[mas/year]}$ & $\mathrm{\mu_\delta\,[mas/year]}$ & Catalogue\\ \hline	& & & & & \\}

\tabletail{\hline \multicolumn{6}{r}{\textit{continued on next page}}\\ \hline}
\tablelasttail{ \hline}

\topcaption{Exoplanet host stars listed in the CCDM, but unlikely a common proper motion pair (same columns as in \ref{tab:systems_cpm}).}
\label{tab:systems_diff_cpm}

\begin{supertabular}{r c c r r r}  
\object{6 Lyn}\,A 	& R	&	-	& $-30.0\pm0.8$ 	& $-338.8\pm0.6$	&	Nomad-1	\\
\object{6 Lyn}\,B 	& -	&	169.91	& $3.6\pm1.6$	 	& $7.2\pm1.6$		&	Nomad-1	\\[0.2cm]

\object{18 Del}\,A 	& R	&	-	& $-47.9\pm0.7$ 	& $-34.3\pm0.3$		&	Nomad-1	\\
\object{18 Del}\,B	& -	&	197.34	& $10.8\pm1.1$ 		& $-12.6\pm1.2$		&	Nomad-1	\\
\object{18 Del}\,C 	& -	&	235.69	& $-6.3\pm1.6$ 		& $-6.4\pm3.2$		&	Nomad-1	\\[0.2cm]

\object{61 Vir}\,A 	& R	&	-	& $-1070.00\pm0.66$ 	& $-1064.22\pm0.49$ 	&	ASCC-2.5 V3	\\
\object{61 Vir}\,B 	& -	&	365.42	& $-31.65\pm3.71$ 	& $-13.56\pm2.26$ 	&	ASCC-2.5 V3	\\[0.2cm]

\object{70 Vir}\,A 	& R	&	 -	& $-234.8\pm0.7$ 	& $-576.1\pm0.5$	&	Nomad-1	\\
\object{70 Vir}\,B 	& -	&	268.29	& $3.9\pm1.3$ 		& $8.7\pm0.9$		& 	Nomad-1	\\[0.2cm]

\object{$\mathrm{\epsilon}$ Tau}\,A 	& R	&	-	& $107.2\pm1.0$	& $-36.7\pm0.8$	&	Nomad-1	\\
\object{$\mathrm{\epsilon}$ Tau}\,B 	& -	&	189.16	& $23.1\pm1.5$ 	& $-18.9\pm1.9$	&	Nomad-1	\\[0.2cm]

\object{$\mathrm{\kappa}$ CrB}\,A 	& 	& -	& $-8.0\pm0.5$ 	& $-347.4\pm0.6$ 	&	Nomad-1	\\
\object{$\mathrm{\kappa}$ CrB}\,B 	& 	& 112.31& $-3.8\pm5.4$ 	& $-20.4\pm5.4$ 	&	Nomad-1		\\[0.2cm]

\object{$\mathrm{\tau}$ Gem}\,A & C	&	-	& $-31.0\pm1.0$ 	& $-48.3\pm0.5$ 	&	Nomad-1	\\
\object{$\mathrm{\tau}$ Gem}\,B & -	&		&  \multicolumn{3}{c}{proper motion not known}	\\
\object{$\mathrm{\tau}$ Gem}\,C & -	&	59.91	& $-64.1\pm9.0$ 	& $-12.5\pm9.0$ 	&	Nomad-1	\\[0.2cm]

\object{HIP 75458}\,A 	& R	&	-	& $-8.2\pm0.3$ 		& $17.3\pm0.4$		&	Nomad-1	\\
\object{HIP 75458}\,B 	& -	&	253.77	& $0.5\pm1.1$ 		& $-5.5\pm1.3$		&	Nomad-1	\\[0.2cm]

\object{HD 33564}\,A 	& R	&	-	& $-79.22\pm0.52$ 	& $161.22\pm0.62$	&	ASCC-2.5 V3	\\
\object{HD 33564}\,B 	& -	&	24.47	& $52.00\pm1.85$	& $-156.6\pm2.99$	&	ASCC-2.5 V3	\\[0.2cm]

\object{HD 62509}\,A 	& R	&	-	& $-625.6\pm1.0$ 	& $-45.9\pm0.5$		&	Nomad-1	\\
\object{HD 62509}\,B 	& -	&	30	& \multicolumn{3}{c}{proper motion not known}	\\
\object{HD 62509 CD} 	& -	&	248.98	& $-4.5\pm1.0$ 		& $-3.7\pm1.3$		&	Nomad-1	\\
\object{HD 62509}\,E 	& -	&	281.26	& $-3.1\pm0.7$ 		& $-14.0\pm0.8$		&	Nomad-1	\\
\object{HD 62509}\,F 	& -	&	304.84	& $-6.5\pm0.7$ 		& $-2.3\pm0.8$		&	Nomad-1	\\
\object{HD 62509}\,G 	& -	&	152.77	& $6.0\pm2.0$ 		& $-4.5\pm1.3$		&	Nomad-1	\\[0.2cm]

\object{HD 81688}\,A 	& R	&	-	& $-6.4\pm0.6$ 		& $-128.1\pm0.5$	&	Nomad-1	\\
\object{HD 81688}\,B	& -	&	71.79	& $-15.0\pm1.0$ 	& $-43.8\pm1.1$		&	Nomad-1	\\
\object{HD 81688}\,C 	& -	&	84.05	& $-17.6\pm1.1$ 	& $-42.6\pm0.7$		&	Nomad-1	\\[0.2cm]

\object{HD 102365}\,A 	& R	&	-	& $-1530.55\pm0.67$	& $402.73\pm0.62$ 	&	ASCC-2.5 V3	\\
\object{HD 102365}\,B 	& -	&	23.95	& $51.6\pm4.4$ 		& $-5.1\pm4.3$ 		&	UCAC-3	\\[0.2cm]

\object{HD 110014}\,A 	& R	&	-	& $-76.90\pm0.54$ 	& $-24.01\pm0.55$	&	ASCC-2.5 V3	\\
\object{HD 110014}\,B	& -	&	176.06	& $-27.19\pm1.82$ 	& $-18.01\pm1.80$	&	ASCC-2.5 V3	\\
\object{HD 110014}\,C 	& -	&	227.06	& $-9.09\pm1.44$ 	& $-1.29\pm2.75$	&	ASCC-2.5 V3	\\
\object{HD 110014}\,D 	& -	&	320.44	& $-26.04\pm1.26$ 	& $-0.94\pm0.64$	&	ASCC-2.5 V3	\\[0.2cm]

\object{HD 121504}\,A 	& R	& 	-	& $-250.5\pm0.7$	& $-84.0\pm0.8$ 	&	Nomad-1	\\
\object{HD 121504}\,B 	& -	&	36.32	& $-15.0\pm1.5$ 	& $2.3\pm1.4$ 		&	Nomad-1	\\[0.2cm]

\object{HD 164922}\,A  & R	&	-	& $389.7\pm0.5$ 	& $-602.4\pm0.5$	&	Nomad-1	\\
\object{HD 164922}\,B  & -	&	96.39	& $6.4\pm5.1$ 		& $-3.9\pm5.1$		&	Nomad-1	\\
\object{HD 164922}\,C 	& -	&	93.09	& $-40.9\pm5.2$ 	& $-56.2\pm4.1$		&	UCAC-3	\\[0.2cm]

\object{HD 192263}\,A 	& R	&	-	& $-63.3\pm1.6$ 	& $262.2\pm0.7$		&	Nomad-1	\\
\object{HD 192263 BC}	& -	&	72.38	& $13.6\pm1.2$ 		& $0.6\pm1.7$		&	Nomad-1	\\
\object{HD 192263}\,D 	& -	&	78.44	& $-4.3\pm5.6$ 		& $-7.9\pm5.6$		&	Nomad-1	\\ 
\end{supertabular}
\end{center}

\vspace{1.5cm}

\begin{table}[h]
\caption[]{Exoplanet host stars with companion candidates, but further epoch observations are needed (same columns as in \ref{tab:systems_cpm}).}
\label{tab:systems_cpm_just_one_epoch}
\begin{center}
\begin{tabular}{r c c r r r}  \hline
System	& EPE planet status	& r [''] 	& $\mathrm{\mu_\alpha \cos{\delta}\,[mas/year]}$ & $\mathrm{\mu_\delta\,[mas/year]}$ & Catalogue\\ \hline
			& 	& 		&		&		&		\\

\object{WASP-2}\,A	& S(R)	&		& $4.9\pm3.9$		& $-50.9\pm6.7$ 	&	UCAC-3\\
\object{WASP-2}\,B& -	&\multicolumn{4}{l}{B is 0.76'' away from A, second epoch needed, see \cite{daemgen_multiplicity_transit_host_stars}} \\[0.2cm]

\object{TrES-2}\,A	& S(R)	&		& $2.89\pm2.50$		& $-3.40\pm2.40$ 	&	ASCC-2.5 V3\\
\object{TrES-2 B}	& -	&\multicolumn{4}{l}{B is 1.09'' away from A, second epoch needed, see \cite{daemgen_multiplicity_transit_host_stars}} \\[0.2cm]

\object{TrES-4}\,A	& S(R)	&		& $-8.09\pm4.80$	& $-33.00\pm4.40$ 	&	ASCC-2.5 V3\\
\object{TrES-4}\,B	& -	&\multicolumn{4}{l}{B is 1.56'' away from A, second epoch needed, see \cite{daemgen_multiplicity_transit_host_stars}} \\ \hline

\end{tabular}
\end{center}
\end{table}

\begin{table}
\caption[]{Extrasolar planets detected with transit or RV observations in closer binaries with a projected stellar separation of $\mathrm{\rho_{app}^\star\leq 1000\,AU}$, sorted by an increasing stellar separation. For the four closest systems a value for the binary \mbox{semi-major} axis ($\mathrm{a_{bin}}$) is known from multi-epoch observations (listed in brackets in the  $\mathrm{\rho_{app}^\star}$ column). If RV and transit measurements are available the true mass of the exoplanet candidate is given in the table.\\[0.3cm] \textcurrency\, ... new system compared to the latest published overview by \citet{mugi_HD125612_HD212301_2009}\\[0.3cm]  \textcolonmonetary\, ... B component is a brown dwarf, see \cite{mugi_HD3651_2006}\\[0.1cm] \textreferencemark\, ... B component is a white dwarf, see \citet{mugi_gl86B_2005}\\[0.1cm] \S\, ... B component is a white dwarf, see \citet{chauvin_2007}\\[0.3cm] \textdied\, ... \citet{reffert_astrom_masses_2011} determined an astrometric mass range for the planet candidate, which is given within the brackets in the planetary mass column}
\label{tab:exo_in_close_binaries}
\begin{center}
\begin{tabular}{c l c c c c l}  \hline
  Note  &	Host star	& $\mathrm{N_{Pl}}$	& $\mathrm{M_{Pl}\sin{i}\;[M_{Jup}]}$ 	& $\mathrm{a_{Pl}\;[AU]}$		& $\mathrm{\rho_{app}^\star\;[AU]}$ 	& Reference\\ \hline

   &			&  		& 			&			& 		& \\

\textdied  &\object{$\mathrm{\gamma}$ Cephei}\,A & 1 & $1.6\;(5\ldots27)$ &$2.04$	& $12.4\,(\mathrm{a_{bin}}=20.2)$	& \cite{campbell_gamma_ceph_1988, gamma_ceph_hatzes_2003}\\
  &			&		&			&			&		& \cite{gamma_ceph_ralph_2007, reffert_astrom_masses_2011}\\[0.1cm]

\textreferencemark  &	\object{Gl 86}\,A & 1 	& $4.01$ 	&$0.11$		& $20.7\,(\mathrm{a_{bin}}=21)$	& \cite{queloz_CORALIE_I_2000, mugi_gl86B_2005} \\[0.1cm]

  &	\object{HD 41004}\,A	& 1 		& $2.54$ 	&$1.64$		& $21.5\,(\mathrm{a_{bin}}=20)$	& \cite{hd41004Ab_2004, raghavan_two_suns_sky_2006}\\[0.1cm]

  &	\object{HD 41004}\,B	& 1 		& $18.40$ 	&$0.02$		& $21.5\,(\mathrm{a_{bin}}=20)$	& \cite{hd41004Bb_2003, raghavan_two_suns_sky_2006}\\[0.1cm]

 &	\object{HD 196885}\,A 	& 1 		& $2.58$ 	&$2.37$		& $23.1\,(\mathrm{a_{bin}}=21)$	& \cite{correia_elodie_IV_2008, fischer_lick_2009}\\
  &			&   	&	&	&	& \cite{chauvin_hd196885B_2011} \\[0.1cm]

  &	\object{$\mathrm{\tau}$ Boo}\,A	& 1 	& $3.90$ 	&$0.05$		& $45.2$	& \cite{butler_tau_boo_1997, raghavan_two_suns_sky_2006}\\[0.1cm]

  &	\object{GJ 3021}\,A	& 1 		& $3.37$ 	&$0.49$		& $68.6$	& \cite{naef_coralie_V_2001, mugi_GJ3021_HD27442_2007}\\[0.3cm]

  &	\object{HD 177830}\,A	& 1 		& $1.28$ 	&$1.00$		& $100.3$	& \cite{vogt_keck_RV_2000, eggenberger_naco_2007}\\
  &			&		&			&			&		& \\[0.1cm]

 &	\object{HD 142}\,A	& 1 		& $1.03$ 	&$1.00$		& $105.0$	& \cite{tinney_hd142_hd23079_2002, raghavan_two_suns_sky_2006}\\[0.1cm]

 &	\object{HD 114762}\,A	& 1 		& $11.02$ 	&$0.30$		& $134.0$	& \cite{latham_HD114762_1989, mugi_2005}\\[0.1cm]

  &	\object{HD 195019}\,A 	& 1 		& $3.70$ 	&$0.14$		& $149.2$	& \cite{fischer_hd195019_hd217107_1999, raghavan_two_suns_sky_2006} \\[0.1cm]

\textcurrency &\object{$\mathrm{\gamma^1}$ Leo}\,A	& 1	&$8.78$ 	&$1.19$		& $165.6$	& \cite{Han_gamma_1_Leo_2010} \\[0.1cm]

 &	\object{HD 189733}\,A	& 1	& $1.13\,(true\,mass)$ 	&$0.03$		& $220.0$	& \cite{bouchy_elodie_II_2005, eggenberger_naco_2007}\\[0.1cm]

 &	\object{HD 16141}\,A	& 1 		& $0.23$ 	&$0.35$		& $222.6$	& \cite{marcy_hd16141_46375_2000, mugi_2005}\\[0.1cm]

\textcurrency  &\object{HD 185269}\,A	& 1 		& $0.94$ 	&$0.08$		& $226.9$	& \cite{johnson_hd185268_2006, ginski_mnras_2012}\\[0.1cm]

 &	\object{HD 212301}\,A	& 1 		& $0.45$ 	&$0.04$		& $233.2$	& \cite{curto_harps_VII_2006, mugi_HD125612_HD212301_2009}\\[0.1cm]

\S &	\object{HD 27442}\,A 	& 1 		& $1.28$ 	&$1.18$		& $238.4$	& \cite{butler_hd160691_hd27442_2001, chauvin_2006} \\
  &			&	&	&	&	&\cite{raghavan_two_suns_sky_2006, mugi_GJ3021_HD27442_2007}\\[0.1cm]

\textcurrency &	\object{HD 28254}\,A	& 1 		& $1.16$ 	&$2.25$		& $241.7$	& \cite{naef_harps_XXIII_2010}\\[0.1cm]

 &	\object{HD 114729}\,A	& 1 		& $0.82$ 	&$2.08$		& $283.5$	& \cite{butler_keck_RV_2003, mugi_2005}\\[0.1cm]

 &	\object{HD 46375}\,A	& 1 		& $0.25$ 	&$0.04$		& $345.7$	& \cite{marcy_hd16141_46375_2000, mugi_calar_alto_survey_2006}\\[0.1cm]

\textcurrency &\object{WASP-8}\,A & 1 		& $2.25\,(true\,mass)$	&$0.08$		& $348.0$	& \cite{queloz_wasp8_2010} \\[0.1cm]

\textcurrency\,,\,\textcolonmonetary &\object{HD 3651}\,A & 1 		& $0.20$ 	&$0.28$		& $478.4$	& \cite{fischer_hd3651_2003, mugi_HD3651_2006} \\[0.1cm]

 &	\object{HD 109749}\,A	& 1 		& $0.28$ 	&$0.06$		& $495.6$	& \cite{fischer_n2k_hd149143_hd109749_2006, desidera_planets_binaries_2007}\\[0.1cm]

  &\object{HD 99492} 		& 1 		& $0.11$ 	&$0.12$		& $589.4$	& \cite{marcy_2005, raghavan_two_suns_sky_2006} \\
  &\,=\,\object{HD 99491}\,B	&	&		&		&		&		\\[0.1cm]

 &	\object{HD 75289}\,A	& 1 		& $0.42$ 	&$0.05$		& $621.4$	& \cite{udry_coralie_II_2000, mugi_HD75289_2004}\\[0.1cm]

 &	\object{HD 188015}\,A	& 1 		& $1.26$ 	&$1.19$		& $676.0$	& \cite{marcy_2005, raghavan_two_suns_sky_2006}\\[0.1cm]

 &\object{$\mathrm{\upsilon}$ And}\,A 	& 3 	& $0.69\ldots 11.6$ &$0.06\ldots 2.55$ & $742.5$	& \cite{butler_upsAnd_1999, lowrance_ups_and_B_2002} \\[0.1cm]

\textcurrency &	\object{GJ 676}\,A	& 1 		& $4.9$ 	&$1.82$		& $788.9$	& \cite{forveille_harps_XXVI_2011}\\[0.1cm]

\textcurrency &	\object{HD 137388}\,A	& 1 		& $0.22$ 	&$0.89$		& $809.4$	& \cite{dumusque_harps_XXX_2011}\\[0.1cm]

 &	\object{HD 142022}\,A	& 1 		& $4.40$ 	&$2.80$		& $890.8$	& \cite{eggenberger_coralie_XIV_hd142022_2006, raghavan_two_suns_sky_2006}\\[0.1cm]

\textcurrency  & \object{11 Com}\,A& 1 		& $19.4$ 	&$1.29$		& $999.0$	& \cite{Liu_11_Comae_2008}\\[0.1cm]

 \hline
\end{tabular}
\end{center}
\end{table}

\begin{table}[t]
\caption[]{Extrasolar planets detected with transit or RV observations in wider binaries  with a projected stellar separation of $\mathrm{\rho_{app}^\star > 1000\,AU}$), sorted by an increasing stellar separation. If RV and transit measurements are available the true mass of the exoplanet candidate is given in the table.\\[0.3cm]  \textcurrency\, ... new system compared to the latest published overview by \citet{mugi_HD125612_HD212301_2009}\\[0.3cm]  \textpilcrow\, ... closer component listed in the CCDM (formerly called B) was disproved by proper motion measurements, but a new wide stellar companion was found and confirmed by common proper motion \citep{mugi_2005}\\[0.1cm] \S\, ... B component is a white dwarf, see \cite{mello_hd147513_WD_1997}}
\label{tab:exo_in_wider_binaries}
\begin{center}
\begin{tabular}{c l c c c c l}  \hline
 Note  & Host star	& $\mathrm{N_{Pl}}$	& $\mathrm{M_{Pl}\sin{i}\;[M_{Jup}]}$ 	& $\mathrm{a_{Pl}\;\;[AU]}$		& $\mathrm{\rho_{app}^\star\;[AU]}$ 	& Reference \\ \hline

   &			&  		& 			&			& 		& \\

 &	\object{HD 11964}\,A	& 2 	& $0.11\,\&\,0.61$	&$0.23\,\&\,3.34$	& $1044$	& \cite{catalogue_nearby_exoplanets, raghavan_two_suns_sky_2006}\\[0.1cm]

 &	\object{55 Cnc}\,A	& 5 	& $0.02\ldots 3.84$ 	&$0.04\ldots 5.77$	& $1053$	& \cite{butler_tau_boo_1997, marcy_55cnc_2002, arthur_HST_rho_Can_2004}\\[0.1cm]

 &	\object{HD 80606}\,A	& 1 	& $3.94\,(true\,mass)$ 	&$0.45$			& $1197$	& \cite{neaf_HD80606b_2001, moutou_HD80606b_2009}\\
  &			&		&			&			&		& \cite{raghavan_two_suns_sky_2006, pont_HD80606b_2009}\\[0.1cm]

\textcurrency &	\object{HD 204941}\,A	& 1 	& $0.27$ 		&$2.56$			& $1447$	& \cite{dumusque_harps_XXX_2011}\\[0.1cm]

 &	\object{HAT-P-1}\,B	& 1 	& $0.52\,(true\,mass)$ 	&$0.06$			& $1557$	& \cite{bakos_hatP1_2007}\\[0.1cm]

 &	\object{HD 101930}\,A	& 1 	& $0.30$ 		&$0.30$			& $2227$	& \cite{lovis_harps_III_2005, mugi_mnras_2007}\\[0.1cm]

\textcurrency &	\object{HD 7449}\,A	& 2 	& $1.11\,\&\,2.00$ 	&$2.30\,\&\,4.96$	& $2348$	& \cite{dumusque_harps_XXX_2011}\\[0.1cm]

 &	\object{HD 89744}\,A	& 1 	& $7.99$ 		&$0.89$			& $2457$	& \cite{korzennik_hd89744_2000, mugi_HD89744_2004}\\[0.1cm]

 &	\object{HD 190360}\,A	& 2 	& $0.06\,\&\,1.50$ 	&$0.13\,\&\,3.92$	& $3293$	& \cite{naef_elodie_II_gj777A_2003, voigt_multip_exoplanets_2005, raghavan_two_suns_sky_2006}\\[0.1cm]

\textpilcrow &	\object{HD 213240}\,A	& 1 	& $4.50$ 		&$2.03$			& $3905$	& \cite{santos_coralie_VI_2001, mugi_2005}\\[0.1cm]

 \S &\object{HD 147513}\,A	& 1	& $1.00$ 		&$1.26$			& $4460$	& \cite{mayor_CORALIE_XII_2004, mugi_gl86B_2005}\\
  &			&		&			&			&		& \cite{mello_hd147513_WD_1997} \\[0.1cm]

 &	\object{HD 222582}\,A	& 1 	& $7.75$ 		&$1.35$			& $4595$	& \cite{vogt_keck_RV_2000, raghavan_two_suns_sky_2006}\\[0.1cm]

\textcurrency  & \object{XO-2}\,A		& 1	& $0.57\,(true\,mass)$	& $0.04$		& $4619$	& \cite{Burke_XO_2b_2007} \\[0.1cm]

 &	\object{HD 125612}\,A	& 3 	& $0.06\ldots7.2$	&$0.05\ldots4.2$	& $4752$	& \cite{fischer_N2K_2007, curto_harps_XXII_2010}\\
  &			&		&			&			&		& \cite{mugi_HD125612_HD212301_2009}\\[0.1cm]

 &	\object{HD 20782}\,A	& 1	& $1.90$ 		&$1.38$			& $9133$	& \cite{jones_angloAustralian_2006, desidera_planets_binaries_2007}\\[0.1cm]

 &	\object{HD 38529}\,A	& 2 	& $0.78\,\&\,17.70$ 	&$0.13\,\&\,3.69$	& $11915$	& \cite{fischer_2001, fischer_2003, raghavan_two_suns_sky_2006}\\ \hline

\end{tabular}
\end{center}
\end{table}

\begin{table}[b]
\caption[]{Extrasolar planets detected with transit or RV observations in stellar systems with more than two components, sorted by the increasing projected separation of the host star and the nearest stellar component ($\mathrm{\rho_{app}^\star}$). For the closest systems a value for the binary \mbox{semi-major} axis ($\mathrm{a_{bin}}$) is known from multi-epoch observations (listed in brackets in the  $\mathrm{\rho_{app}^\star}$ column). If RV and transit measurements are available the true mass of the exoplanet candidate is given in the table.\\[0.3cm] \textcurrency\, ... new system compared to the latest published overview by \citet{mugi_HD125612_HD212301_2009}}
\label{tab:exo_in_triples}
\begin{center}
\begin{tabular}{c l c c c c c l}\hline
Note   &	Host-star	&  $\mathrm{N_{Pl}}$	& $\mathrm{N_{\star}}$	& $\mathrm{M_{Pl}\sin{i}\;[M_{Jup}]}$ 	& $\mathrm{a_{Pl}\;[AU]}$		& $\mathrm{\rho_{app}^\star\;[AU]}$ 	& Reference \\ \hline

   &			&  & 		& 		&		& 		& \\

\textcurrency  &	\object{HD 126614}\,A	& 1 & 3		& $0.38$ 	&$2.35$		& $36.2$	& \cite{howard_Cal_Planet_search_2010, ginski_mnras_2012}\\[0.1cm]

  &	\object{HD 19994}\,A	& 1 & 3		& $1.68$ 	&$1.42$		& $51.5$	& \cite{mayor_CORALIE_XII_2004, raghavan_two_suns_sky_2006}\\
   &			&   &		&	&	& 	&\cite{roell_paris_2011} \\
   &			&   &		&		&		&		& \\[0.2cm]

\textcurrency    &	\object{GJ 667}\,C	& 2 & 3	 & $0.018\,\&\,0.014$ &$0.05\,\&\,0.12$	& $227.0$	& \cite{escude_gj667C_2012}\\[0.1cm]

  &	\object{HD 65216}\,A	& 1 & 3		& $1.21$ 	&$1.37$		& $256.3$	& \cite{mayor_CORALIE_XII_2004, mugi_mnras_2007}\\[0.1cm]

\textcurrency  &	\object{HD 132563}\,B	& 1 & 3		& $1.49$ 	&$2.62$		& $365.2$	& \cite{desidera_hd132563_2011}\\[0.1cm]

  &	\object{HD 196050}\,A	& 1 & 3		& $3.00$ 	&$2.50$		& $511.2$	& \cite{jones_angloAustralian_2002, mugi_2005}\\
  &			&   &		&			&			&		& \cite{eggenberger_naco_2007} \\[0.1cm]

 &	\object{HD 178911}\,B	& 1 & 3		& $6.29$ 	&$0.32$		& $794.3$	& \cite{zucker_hd178911_2002, eggenberger_2003} \\[0.1cm]

 &	\object{16 Cyg}\,B	& 1 & 3		& $1.68$ 	&$1.68$		& $859.7$	& \cite{cochran_16CygniB_1997, raghavan_two_suns_sky_2006} \\
   &			&   &		&		&		&		& \\[0.2cm]

\textcurrency  &	\object{30 Ari}\,B	& 1 & 3		& $9.88$ 	&$0.995$	& $1517$	& \cite{eike_30_Ari_2009}\\[0.1cm]

 &	\object{HD 40979}\,A	& 1 & 3		& $3.32$ 	&$0.81$		& $6395$	& \cite{fischer_2003, eggenberger_2003}\\
 &			&   &		&			&			&		& \cite{mugi_GJ3021_HD27442_2007}\\

 \hline
\end{tabular}
\end{center}
\end{table}

\end{appendix}

%
\end{document}